\documentclass[twocolumn,showpacs,amsmath,amssymb,prb]{revtex4}

\usepackage[dvips]{graphicx}
\usepackage{dcolumn}
\usepackage{braket}
\usepackage{bm}
\usepackage{txfonts}

\begin{document}

\title{Strong-Coupling Superconductivity with Mixed Even- and Odd-Frequency Pairing}
\author{Hiroaki Kusunose}
\affiliation{Department of Physics, Ehime University, Matsuyama, Ehime 790-8577, Japan}
\author{Masashige Matsumoto}
\affiliation{Department of Physics, Faculty of Science, Shizuoka University, Shizuoka 422-8529, Japan}
\author{Mikito Koga}
\affiliation{Department of Physics, Faculty of Education, Shizuoka University, Shizuoka 422-8529, Japan}

\date{\today}

\begin{abstract}
We investigate general structure of Landau free energy for stabilizing a novel superconducting state with both even-frequency and odd-frequency components in gap function.
On the basis of the Luttinger-Ward functional, we elucidate an emergent mixing between different ``parity'' in time.
The simplest case of the conventional $s$-wave singlet mixed with the odd-frequency triplet state under broken time-reversal symmetry is examined to demonstrate the anomalous structure of the free-energy functional.
The induced odd-frequency component alters behaviors of physical quantities from those obtained by neglecting the odd-frequency component.
The novel mixed state may also be relevant to strong-coupling superconductivity coexisting with ferromagnetism.
\end{abstract}

\maketitle

\section{Introduction}

A symmetry of order parameter is one of central issues in study of superconductivity.
The isotropic $s$-wave singlet pairing in the original BCS theory has been extended widely to anisotropic ones including spin-triplet states in the strongly correlated systems\cite{RDS12}.
Alternative direction of extension was proposed by Berezinskii\cite{Berezinskii74}, where the gap function is ``anisotropic'' in time, i.e., with an odd-frequency dependence.
This novel state has been investigated extensively in the context of the disordered Fermi liquid\cite{Kirkpatrick91,Belitz92}, the high-$T_{c}$ cuprates\cite{Balatsky92,Abrahams93,Abrahams95}, the doped triangular antiferromagnets\cite{Vojta99}, one-dimensional organic systems\cite{Shigeta09,Shigeta11}, quantum critical spin fluctuations in heavy fermions\cite{Fuseya03}, strong-coupling local electron-phonon systems\cite{Kusunose11}, and orbitally degenerate systems with full spin polarization\cite{Hotta09}.

Besides exploration of possible odd-frequency pairings in homogeneous bulk systems, it has been argued that the odd-frequency pair amplitude arises in the spatially nonuniform situation, particularly at the surface/interface of the sample, quite ubiquitously\cite{Bergeret05,Tanaka12}.
It is more promising to induce the odd-frequency pair amplitude in the nonuniform system with lower symmetry.
The case of spatially lower symmetry has also been investigated in noncentrosymmetric systems without inversion symmetry, where even- and odd-parity components are mixed, and it influences many of superconducting properties in a profound way\cite{Bauer12}.
Quite analogously a mixing of even- and odd-frequency components should occur in principle, when the time-reversal symmetry is broken.
The present authors proposed the mixing of the conventional $s$-wave singlet with the odd-frequency triplet component under applied uniform magnetic fields through Zeeman splitting of the conduction band\cite{Matsumoto12}.
Such a situation may also be relevant to superconductivity coexisting with ferromagnetism\cite{Hotta09,Aoki12}.

When the odd-frequency component is involved, a delicate treatment of the superconducting state should be required.
Namely, in describing a superconducting state, there exist two independent gap functions, $\Delta({\bm k},i\omega_{n})$ and its particle-hole converted counterpart $\Delta^{+}({\bm k},i\omega_{n})$, which are usually considered as hermite conjugate pairs.
However, it has been argued that the relation between them is nontrivial in the case of the odd-frequency pairing\cite{Kusunose11a,Belitz99}. 
A guiding principle of determining the relation between them is real and minimum condition of the free energy, and it is concluded that the correct choice is $\Delta^{+}({\bm k},i\omega_{n})=\Delta({\bm k},i\omega_{n})^{*}$\cite{Kusunose11a,Belitz99}, although it has long been believed $\Delta^{+}({\bm k},i\omega_{n})=-\Delta({\bm k},i\omega_{n})^{*}$ in the case of the odd-frequency pairing.
The latter relation was the main source of fatal deficiency of the bulk odd-frequency pairing, such as thermodynamic instability and unphysical negative Meissner kernel in the superconducting phase\cite{Kusunose11a,Solenov09}.

Nevertheless, we shall discuss in this paper that the above relation, $\Delta^{+}({\bm k},i\omega_{n})=\Delta({\bm k},i\omega_{n})^{*}$, is no longer satisfied simultaneously for both components in a coexistence case of the even-frequency and the odd-frequency pairings\cite{Matsumoto12}.
Consequently, the {\it majority} component of them takes the ordinary sign, $\Delta^{+}({\bm k},i\omega_{n})=+\Delta({\bm k},i\omega_{n})^{*}$, while the {\it minority} component must have the {\it opposite} sign, $\Delta^{+}({\bm k},i\omega_{n})=-\Delta({\bm k},i\omega_{n})^{*}$, in order to gain interference energy between two components.
As a result, it exhibits different behaviors in various physical quantities from those obtained by neglecting this minority component with the anomalous negative sign relation.

The paper is organized as follows.
In \S2, we discuss general structure of the free-energy functional in the coexistence case on the basis of the Luttinger-Ward functional theory.
The real and minimum condition of the free energy requires the opposite sign relations for spin singlet and triplet components.
In \S3, we examine the $s$-wave singlet with the induced odd-frequency triplet under the applied magnetic fields, based on the ordinary electron-phonon model with a single Einstein frequency.
The anomalous feature of the free-energy functional is demonstrated in the strong-coupling limit with single gap approximation.
We show the comparisons of the specific heat and the superfluid density with or without the odd-frequency component by solving the full Eliashberg equations.
The last section summarizes the paper.
Throughout the paper, we have neglected the orbital effect of the magnetic fields.
The derivation of the explicit form of the free-energy functional is given in Appendix A.
The noncentrosymmetric case with the Rashba spin-orbit coupling is briefly discussed in Appendix B.

\section{General argument on even-frequency and odd-frequency mixing}

\subsection{Free energy in terms of Luttinger-Ward functional}

Let us begin with the Luttinger-Ward functional for the thermodynamic potential\cite{Luttinger60,AGD} measured from non-interacting one,
\begin{multline}
\Omega[\hat{G}]=-\frac{T}{2}\sum_{k}{\rm Tr}\,\left[
\ln\left\{-\hat{G}^{-1}(k)\right\}-\left\{\hat{1}-\hat{G}_{0}^{-1}(k)\hat{G}(k)\right\}
\right]
\\
+\frac{\Phi[\hat{G}]}{2}-\Omega_{0},
\end{multline}
where $k=({\bm k},i\omega_{n})$ is the 4-dimensional momentum with the fermionic Matsubara frequency, $\omega_{n}=(2n+1)\pi T$ at temperature, $T$.
The trace is taken over the $4\times4$ Nambu space defined as $\Psi^{\dagger}(k)=[c_{\uparrow}^{\dagger}(k),\,c_{\downarrow}^{\dagger}(k),\,c_{\uparrow}^{}(-k),\,c_{\downarrow}^{}(-k)]$ (a hat represents a $4\times4$ matrix).
$\hat{G}_{0}(k)$ and $\Omega_{0}=-(T/2)\sum_{k}{\rm Tr}\ln(-\hat{G}_{0}^{-1}(k))$ are the non-interacting Green's function and its free energy (thermodynamic potential), respectively.
Here, the thermodynamic potential is expressed in terms of the full Green's function, $\hat{G}(k)$.
It can be shown to be stationary at the physical $\hat{G}(k)$ that satisfies the Dyson equation, $\hat{G}^{-1}(k)=\hat{G}_{0}^{-1}(k)-\hat{\Sigma}(k)$, provided that the self energy is given by the functional derivative as $\Sigma_{ij}(k)=T^{-1}\delta\Phi/\delta G_{ji}(k)$.
The functional $\Phi$ consists of so called the skeleton diagram of $\hat{G}(k)$ with interaction lines.

To be specific, we consider,
\begin{equation}
\Phi[\hat{G}]=\frac{T^{2}}{2}\sum_{kk'}V(k-k'){\rm Tr}\left[\hat{G}(k)\hat{\rho}_{3}\hat{G}(k')\hat{\rho}_{3}\right],
\label{lwfel}
\end{equation}
where $\hat{\rho}_{3}$ is the $z$-component of the Pauli matrix acting on the particle-hole space, and $V(q)$ is an arbitrary $q$-dependent scalar interaction ($V>0$ represents an attraction), which is a real and even function of $q$.
It can easily be shown that the stationary condition gives the Eliashberg equation\cite{Eliashberg60} with the self energy,
\begin{equation}
\hat{\Sigma}(k)=T\sum_{k'}V(k-k')\hat{\rho}_{3}\hat{G}(k')\hat{\rho}_{3}.
\end{equation}

In order to perform the following Landau free-energy argument, it is more convenient to transform $\Omega[\hat{G}]$ to a functional with respect to the self energy, $\Omega[\hat{\Sigma}]$.
It can be accomplished by the Legendre transformation\cite{Potthoff03}, $\Phi[\hat{G}]=\Theta[\hat{\Sigma}]+T\sum_{k}{\rm Tr}[\hat{G}(k)\hat{\Sigma}(k)]$ as
\begin{equation}
\Omega[\hat{\Sigma}]=-\frac{T}{2}\sum_{k}{\rm Tr}\,\ln\left[-\left\{\hat{G}_{0}^{-1}(k)-\hat{\Sigma}(k)\right\}\right]+\frac{\Theta[\hat{\Sigma}]}{2}-\Omega_{0},
\label{glf}
\end{equation}
with $G_{ij}(k)=-T^{-1}\delta\Theta/\delta\Sigma_{ji}(k)$.
Equation (\ref{lwfel}) corresponds to
\begin{equation}
\Theta[\hat{\Sigma}]=-\frac{1}{2}\sum_{kk'}W(k-k'){\rm Tr}\left[\hat{\Sigma}(k)\hat{\rho}_{3}\hat{\Sigma}(k')\hat{\rho}_{3}\right],
\end{equation}
where $W(k-k')=\sum_{x}V^{-1}(x)e^{-i(k-k')x}$ is the matrix inverse of $V(k-k')$ in the $(k,k')$ space ($W>0$ represents an attraction).
The stationary condition for $\Omega[\hat{\Sigma}]$ again gives the Dyson (Eliashberg) equation.
Since the anomalous component of the self-energy matrix is the superconducting order parameter, (\ref{glf}) is regarded as the Landau free-energy functional for superconductivity.
It is straightforward to extend the present formulation for non-scalar general interaction by modifying only the part of $\Theta[\hat{\Sigma}]$ in the free-energy functional, $\Omega[\hat{\Sigma}]$.

Once the stationary solution of (\ref{glf}) is obtained, the physical equilibrium free energy is given by
\begin{equation}
\Omega_{\rm s}=-\frac{T}{2}\sum_{k}{\rm Tr}\left[\ln\left\{-\hat{G}^{-1}(k)\right\}+\frac{1}{2}\hat{G}(k)\hat{\Sigma}(k)\right]-\Omega_{0},
\end{equation}
in which the Dyson equation, $\hat{G}^{-1}(k)=\hat{G}_{0}^{-1}(k)-\hat{\Sigma}(k)$, is satisfied.
It is useful to note that the derivative of $\Omega_{\rm s}$ with respect to a parameter $x$ can be obtained by the explicit derivative of $\Omega$ evaluated at the stationary, i.e., $d\Omega_{\rm s}/dx=\partial\Omega/\partial x|_{\hat{\Sigma}_{\rm s}}$.

\subsection{Decomposition of singlet and triplet components}

Let us introduce the components of the $4\times4$ matrix as follows,
\begin{align}
&
\hat{G}_{0}^{-1}(k)-\hat{\Sigma}(k)=
\begin{pmatrix}
z(k) & -\Delta(k) \\
-\Delta^{+}(k) & z^{+}(k)
\end{pmatrix},
\notag\\&\quad\quad
z(k)=z_{0}(k)\sigma_{0}+{\bm z}(k)\cdot{\bm\sigma}=-z^{+}(-k)^{T},
\notag\\&\quad\quad\quad
z_{0}(k)=i\omega_{n}-\xi_{\bm k}-\Sigma_{0}(k),
\quad
{\bm z}(k)={\bm h}-{\bm\Sigma}(k),
\notag\\&\quad\quad
\Delta(k)=d_{0}(k)\tau_{0}+{\bm d}(k)\cdot{\bm\tau}=-\Delta(-k)^{T},
\notag\\&\quad\quad
\Delta^{+}(k)=d_{0}^{+}(k)\tau_{0}^{\dagger}+{\bm d}^{+}(k)\cdot{\bm\tau}^{\dagger}=-\Delta^{+}(-k)^{T},
\\
&\hat{G}(k)=-\Braket{T_{\tau}\Psi(k)\Psi^{\dagger}(k)}
=\begin{pmatrix}
G(k) & -F(k) \\
-F^{+}(k) & G^{+}(k)
\end{pmatrix},
\notag\\&\quad\quad
G(k)=g_{0}(k)\sigma_{0}+{\bm g}(k)\cdot{\bm\sigma}=-G^{+}(-k)^{T},
\notag\\&\quad\quad
F(k)=f_{0}(k)\tau_{0}+{\bm f}(k)\cdot{\bm\tau}=-F(-k)^{T},
\notag\\&\quad\quad
F^{+}(k)=f_{0}^{+}(k)\tau_{0}^{\dagger}+{\bm f}^{+}(k)\cdot{\bm\tau}^{\dagger}=-F^{+}(-k)^{T},
\end{align}
where the superscripts $\dagger$ and $T$ represent the hermite conjugate and the transpose of a $2\times2$ matrix, and $\xi_{\bm k}$ and ${\bm h}$ are the one-particle energy measured from the chemical potential and the external magnetic field, respectively.
The normal part is decomposed into the charge and the spin components by $\sigma_{0}$ and ${\bm\sigma}$, which are the $2\times2$ unit matrix and the vector of the Pauli matrix acting on the spin space.
The relations, $z(k)=-z^{+}(-k)^{T}$ and $G(k)=-G^{+}(-k)^{T}$ are obtained by definition of the Green's functions.
The anomalous self energy is also decomposed into the singlet and the triplet components by $\tau_{0}=i\sigma_{0}\sigma_{2}$ and ${\bm\tau}=i{\bm\sigma}\sigma_{2}$.
The anti-commutation relation of fermions requires the even property, $d_{0}(k)=d_{0}(-k)$ and $d_{0}^{+}(k)=d_{0}^{+}(-k)$ in the singlet channel, and the odd property, ${\bm d}(k)=-{\bm d}(-k)$ and ${\bm d}^{+}(k)=-{\bm d}^{+}(-k)$ in the triplet channel.
In this paper, we assume the presence of the inversion symmetry, and $z^{+}(k)=-z(k)^{*}$, i.e., $z^{+}_{\alpha}(k)=-z_{\alpha}(k)^{*}$ ($\alpha=0,1,2,3$) holds.
Note that the $d_{\alpha}^{+}(k)$ [$f_{\alpha}^{+}(k)$] are independent of $d_{\alpha}(k)$ [$f_{\alpha}(k)$], and the relation between them are determined by real and minimum condition of the free-energy functional as follows.

With this preliminary, we expand the free-energy functional (\ref{glf}) with respect to $d_{0}(k)$, ${\bm d}(k)$, $d_{0}^{+}(k)$, and ${\bm d}^{+}(k)$.
After some manipulation (see, Appendix in detail), we obtain the lowest-order expression of the Landau expansion as
\begin{align}
&\Delta\Omega_{2}[d_{\alpha},d_{\alpha}^{+}]=\sum_{kk'}W(k-k')\left[d_{0}(k)d_{0}^{+}(k')+{\bm d}(k)\cdot{\bm d}^{+}(k')\right]
\notag\\&\quad
+T\sum_{k}
\frac{1}{|w|^{2}}\biggl[
(z_{0}z_{0}^{+}-{\bm z}\cdot{\bm z}^{+})(d_{0}d_{0}^{+}+{\bm d}\cdot{\bm d}^{+})
\notag\\&\quad\quad\quad\quad
+({\bm z}\times{\bm d})\cdot({\bm z}^{+}\times{\bm d}^{+})+({\bm z}\times{\bm d}^{+})\cdot({\bm z}^{+}\times{\bm d})
\notag\\&\quad\quad\quad\quad
-{\bm n}_{0}\cdot {\bm m}_{0}
+{\bm n}_{1}\cdot{\bm m}_{1}
+{\bm n}_{2}\cdot{\bm m}_{2}
\biggr],
\label{glex2}
\end{align}
where the free-energy functional is measured from that of the normal state, and $w=|z|$, $z_{\alpha}$ and $z^{+}_{\alpha}$ are evaluated in the normal state.
The symmetric and the antisymmetric vectors have been introduced as
\begin{align}
&{\bm n}_{0}(k)=z_{0}(k){\bm z}^{+}(k)+z_{0}^{+}(k){\bm z}(k)={\bm n}_{0}(-k),
\notag\\
&{\bm n}_{1}(k)=i[{\bm z}(k)\times{\bm z}^{+}(k)]=-{\bm n}_{1}(-k),
\notag\\
&{\bm n}_{2}(k)=z_{0}(k){\bm z}^{+}(k)-z_{0}^{+}(k){\bm z}(k)=-{\bm n}_{2}(-k).
\label{nvec}
\end{align}
Due to the relation, $z^{+}_{\alpha}(k)=-z_{\alpha}(k)^{*}$, ${\bm n}_{0}(k)$ and ${\bm n}_{1}(k)$ are real, while ${\bm n}_{2}(k)$ is pure imaginary.
The corresponding symmetric and antisymmetric vectors composed of $d_{\alpha}(k)$ and $d_{\alpha}^{+}(k)$ are given by
\begin{align}
&{\bm m}_{0}(k)=i[{\bm d}(k)\times{\bm d}^{+}(k)]={\bm m}_{0}(-k),
\notag\\
&{\bm m}_{1}(k)=d_{0}^{+}(k){\bm d}(k)-d_{0}(k){\bm d}^{+}(k)=-{\bm m}_{1}(-k),
\notag\\
&{\bm m}_{2}(k)=d_{0}^{+}(k){\bm d}(k)+d_{0}(k){\bm d}^{+}(k)=-{\bm m}_{2}(-k).
\label{mvec}
\end{align}

The explicit expressions of ${\bm n}_{i}(k)$ are given by
\begin{align}
&\frac{1}{2}{\bm n}_{0}(k)=\xi{\bm h}+\Sigma_{0}'{\bm h}-\xi{\bm\Sigma}'+\omega_{n}{\bm\Sigma}''-(\Sigma_{0}{\bm\Sigma}^{*})',
\\
&{\bm n}_{1}(k)=2({\bm h}\times{\bm\Sigma}'')-i({\bm\Sigma}\times{\bm\Sigma}^{*}),
\\
&\frac{1}{2i}{\bm n}_{2}(k)= -\omega_{n}{\bm h}+\Sigma_{0}''{\bm h}+\xi{\bm\Sigma}''+\omega_{n}{\bm\Sigma}'-(\Sigma_{0}{\bm\Sigma}^{*})'',
\end{align}
where the prime and the double prime mean the real and the imaginary part, respectively.
Since the normal self energy does not change the symmetry of the system in the normal phase (we do not consider a spontaneous symmetry breaking in the normal self energy), we neglect the normal self energy in considering symmetry properties of ${\bm n}_{i}(k)$.
Then, ${\bm n}_{0}\sim \xi{\bm h}$, ${\bm n}_{1}\sim0$, ${\bm n}_{2}\sim i\omega_{n}{\bm h}$, and if one of them is finite, a mixing between different components of $d_{\alpha}$ arises.

Let us first consider the case ${\bm h}=0$, i.e., ${\bm n}_{i}(k)=0$.
In this case, either the pure singlet or the triplet pairing is realized depending on the structure of the attraction.
Since the equilibrium free energy must be a real quantity, a stationary solution must have the relation, $d_{\alpha}^{+}(k)=\phi_{\alpha} d_{\alpha}(k)^{*}$ with $\phi_{\alpha}$ taking either $+1$ or $-1$.
Note that $\phi_{\alpha}$ should be independent of $k$, since $d_{\alpha}(k)$ and $d_{\alpha}^{+}(k)$ are mutually equal counterparts.
For a pure singlet or a triplet state, it is the sufficient condition for the real free energy.
Then, the sign $\phi_{\alpha}=+1$ should be chosen by the minimum condition of the free-energy functional through a phase transition to the superconducting state.
Consequently, $d_{\alpha}^{+}(k)=d_{\alpha}(k)^{*}$ always holds for the pure singlet or the triplet pairing, irrespective of its symmetry in the frequency domain, as was discussed previously\cite{Kusunose11a}.
Hereafter, the free-energy functional, that is defined over the hyperplane under the constraint between $d_{\alpha}(k)$ and $d_{\alpha}^{+}(k)$, is referred as the {\it constrained} free-energy functional.
Note that the constrained free-energy functional is always real by definition.

Next, we elucidate the effect of the external magnetic field, ${\bm h}$.
In this case, ${\rm Re}({\bm n}_{0})$ and ${\rm Im}({\bm n}_{2})$ become finite (${\bm n}_{0}=0$ in the presence of the particle-hole symmetry).
Thus, the terms coupled with ${\bm n}_{0}$ and ${\bm n}_{2}$ in the free-energy functional induce ${\bm m}_{0}$ and ${\bm m}_{2}$.
Due to the real condition of the free energy, ${\bm m}_{0}$ and ${\bm m}_{2}$ must be real and pure imaginary, respectively.
In order to satisfy these conditions, the singlet and the triplet components have the opposite sign relation, i.e.,
\begin{equation}
d_{0}^{+}(k)=\phi d_{0}(k)^{*},
\quad
{\bm d}^{+}(k)=-\phi{\bm d}(k)^{*},
\quad (\phi=+1,\,{\rm or}\,-1).
\label{ddpcond}
\end{equation}
It should be noted that ${\bm n}_{2}$ does not mix the spatial parity, but mix the even-frequency and the odd-frequency components.
Namely, either the singlet or the triplet has the odd-frequency dependence.
By inserting (\ref{ddpcond}) into (\ref{mvec}), it is easily shown that the relative phase factor between the singlet and the triplet components is pure imaginary.

Using (\ref{ddpcond}), the explicit form of the equilibrium free energy is eventually given by
\begin{multline}
\Omega_{\rm s}=-\frac{T}{2}\sum_{k}\ln X(k)-\frac{T}{2}\sum_{k}\biggl[g_{0}(k)\Sigma_{0}(k)+{\bm g}(k)\cdot{\bm\Sigma}(k)
\\
-\phi\biggl\{f_{0}(k)d_{0}(k)^{*}-{\bm f}(k)\cdot{\bm d}(k)^{*}\biggr\}
\biggr]-\Omega_{0},
\label{frees}
\end{multline}
where $g_{\alpha}(k)$, $\Sigma_{\alpha}(k)$, $f_{\alpha}(k)$ and $d_{\alpha}(k)$ satisfy the Eliashberg equations, (\ref{ee1}), (\ref{ee2}) with (\ref{gf1}) and (\ref{gf2}).
The explicit expression of $X(k)=|-\hat{G}^{-1}(k)|$ is given by (\ref{exdet}).
The anomalous Green's functions satisfy the relations similar to (\ref{ddpcond}) as
\begin{equation}
f_{0}^{+}(k)=\phi f_{0}(k)^{*},
\quad
{\bm f}^{+}(k)=-\phi{\bm f}(k)^{*}.
\label{frel}
\end{equation}

The sign of $\phi$ is determined by the minimum condition of the constrained free-energy functional, or equivalently the solvability of the Eliashberg equations in the superconducting state.
As will be shown explicitly in the next section, a stable solution of the Eliashberg equations can be found, only when we use the correct choice of the sign, $\phi$.
Note also that this stable solution corresponds to the {\it saddle point} of the constrained free-energy functional.
An arbitrary initial condition always converges to this saddle point in the iterative procedure, as long as the correct choice of $\phi$ is used.
In a coexistence case and one component dominates over other components, the majority component would determine the sign of $\phi$, i.e., $\phi=+1$ for the singlet majority, while $\phi=-1$ for the triplet majority.
In the competing case, however, the comparison of the equilibrium free energies both for $\phi=\pm1$ is necessary to determine the correct sign of $\phi$.

It is important to note that the contributions to the equilibrium free energy from the singlet and the triplet components have opposite signs as in the last term of (\ref{frees}), as a consequence of (\ref{ddpcond}).
In other words, the mixing of the singlet and the triplet pairings tends to increase the net free energy as compared with the case of neglecting the odd-frequency component.
As a result, it alters behaviors of various physical quantities such as the reduction of $T_{c}$.
The explicit examples are given in the next section.

\section{The $s$-wave singlet with odd-frequency triplet state under magnetic fields}

In order to elucidate essential features of the even- and the odd-frequency mixing, we consider the $s$-wave singlet and triplet pairings under magnetic field $h$ along $z$ axis as the simplest example.
The orbital effect is not taken into account for simplicity, so that the discussions can be applied in low fields.

\subsection{The formulation based on the free-energy functional}

Since $n_{2z}\sim i\omega_{n}h\ne0$ in this case, the singlet ($d_{0}$) and the $S_{z}=0$ triplet ($d_{3}$) components are mixed\cite{notes1}, and the relative phase becomes pure imaginary as noted in the previous section.
We assume the particle-hole symmetry after performing the momentum integration, the ${\bm n}_{0}$ term in the free-energy functional vanishes.
By these reasons, we denote $d_{0}(i\omega_{n})$ and $d_{3}(i\omega_{n})$ as $d_{{\rm s}n}$ and $id_{{\rm t}n}$, then both $d_{{\rm s}n}$ and $d_{{\rm t}n}$ can be chosen as real.
The anti-commutation relation requires that $d_{{\rm s}n}$ and $d_{{\rm t}n}$ have the even-frequency and the odd-frequency dependences, respectively.
Therefore, without considering the retardation effect (neglecting $\omega_{n}$ dependence), $d_{{\rm t}n}$ should vanish as in past studies.

The normal part of the self energies $\Sigma_{0}(i\omega_{n})$ and $\Sigma_{3}(i\omega_{n})$ are finite as well.
It is shown that $\Sigma_{0}(i\omega_{n})$ [$\Sigma_{3}(i\omega_{n})$] is pure imaginary odd [real even] function, so that we denote $\Sigma_{0}(i\omega_{n})=i\omega_{n}(1-Z_{n})$ and $\Sigma_{3}(i\omega_{n})=\Sigma_{3n}$.
Then, both $Z_{n}$ and $\Sigma_{3n}$ are the real even functions.
$Z_{n}$ is so-called the mass enhancement factor.

For notational simplicity, we introduce $\Delta_{n}=d_{{\rm s}n}+id_{{\rm t}n}=\Delta_{-n-1}^{*}=\phi \Delta_{n}^{+}$ corresponding to the $(\uparrow,\downarrow)$ component, and $\Sigma_{n}=i\omega_{n}(1-Z_{n})+\Sigma_{3n}=\Sigma_{-n-1}^{*}$ corresponding to the $(\uparrow,\uparrow)$ component.
We adopt the local electron-phonon-type attraction with retardation,
\begin{equation}
v_{m,n}=\frac{\rho_{\rm F}V_{m,n}}{\lambda}=(W^{-1})_{m,n}\equiv\frac{\omega_{\rm E}^{2}}{\omega_{\rm E}^{2}+(\omega_{m}-\omega_{n})^{2}},
\label{epint}
\end{equation}
where $\omega_{\rm E}$ is the characteristic range of the interaction, $\rho_{\rm F}$ is the density of states per spin at the Fermi energy, and $\lambda>0$ is the dimensionless coupling constant for the attraction.

The Einstein-phonon attraction (\ref{epint}) is the same order of magnitude $\lambda$ both in the even-frequency (the even part with respect to $\omega_{n}\to-\omega_{n}$) and the odd-frequency (the odd part  with respect to $\omega_{n}\to-\omega_{n}$) channels (hence the same order of $T_{c}$), unless the normal self-energy effect (mass enhancement) is not taken into account\cite{Kusunose11}.
Owing to the odd property of the attraction in the odd-frequency channel, which is vanishing toward $T\to0$, the odd-frequency pairing is suppressed at low temperatures, yielding the reentrant behavior in the case of the pure odd-frequency pairing.
In addition to this, the mass enhancement significantly suppresses the odd-frequency pairing, so that it is naturally expected that the induced odd-frequency component discussed below is also small as compared with the majority even-frequency component\cite{Matsumoto12}.

Performing the ${\bm k}$ integration, we obtain the free-energy functional as
\begin{align}
&\Omega[\hat{\Sigma}]=
\frac{\rho_{\rm F}}{\lambda}\sum_{mn}W_{m,n}\left(\phi \Delta_{m}\Delta_{n}-\Sigma_{m}\Sigma_{n}\right)
-\frac{T}{2}\sum_{n}\sum_{{\bm k}}\ln\,X(k)-\Omega_{0}
\notag\\&\quad\quad
=\frac{\rho_{\rm F}}{\lambda}\sum_{mn}W_{m,n}\left(\phi \Delta_{m}\Delta_{n}-\Sigma_{m}\Sigma_{n}\right)
-2\rho_{\rm F}\pi T\sum_{n}(D_{n}-|\omega_{n}|),
\end{align}
where $X(k)=|\xi_{\bm k}^{2}+D_{n}^{2}|^{2}$ with
\begin{equation}
D_{n}=\sqrt{(\omega_{n}+i\Sigma_{n}-ih)^{2}+\phi \Delta_{n}^{2}}=D_{-n-1}^{*}.
\end{equation}

The stationary condition, $\delta\Omega/\delta \Sigma_{n}=\delta\Omega/\delta \Delta_{n}=0$ leads to the Eliashberg equations,
\begin{align}
&\Sigma_{m}=\lambda\pi T\sum_{n}v_{m,n}G_{n},
\notag\\
&\Delta_{m}=\lambda\pi T\sum_{n}v_{m,n}F_{n},
\label{eeqs}
\end{align}
where the local Green's functions are defined as
\begin{align}
&G_{n}\equiv \frac{-1}{2\pi\rho_{\rm F}}\sum_{{\bm k}}\frac{\partial}{\partial \Sigma_{n}}\ln\,X(k)=-i\frac{\omega_{n}+i\Sigma_{n}-ih}{D_{n}}=G_{-n-1}^{*},
\\
&F_{n}\equiv \frac{\phi}{2\pi\rho_{\rm F}}\sum_{{\bm k}}\frac{\partial}{\partial \Delta_{n}}\ln\,X(k)=\frac{\Delta_{n}}{D_{n}}=F_{-n-1}^{*}.
\end{align}
These Eliashberg equations are examined numerically in the previous work\cite{Matsumoto12}.
Using the solution of the Eliashberg equations, the equilibrium free-energy difference is expressed as
\begin{multline}
\Delta \Omega_{\rm s}=2\rho_{\rm F}\pi T\sum_{n=0}^{\infty}{\rm Re}\left[
\left(\phi F_{n}\Delta_{n}-G_{n}\Sigma_{n}\right)+\omega_{n}(1-Z_{{\rm N}n})
\right.\notag\\\left.
-2(D_{n}-Z_{{\rm N}n}\omega_{n})
\right],
\end{multline}
where $Z_{{\rm N}m}=1+(\lambda\pi T/|\omega_{m}|)\sum_{n=0}^{\infty}(v_{m,n}-v_{m,-n-1})$ is the mass-enhancement factor in the normal state.
By appropriate numerical differentiation of $\Delta\Omega_{\rm s}$, we obtain the thermodynamic quantities such as the specific heat.
The superfluid density may be obtained by the standard derivation as\cite{AGD,Kusunose11a}
\begin{align}
n_{\rm s}&=2\pi T\sum_{n=0}^{\infty}{\rm Re}\left(\frac{\Delta_{n}^{2}}{D_{n}^{3}}\right)\phi
\notag\\&
=2\pi T\sum_{n=0}^{\infty}\left[{\rm Re}\left(\frac{1}{D_{n}^{3}}\right)(d_{{\rm s}n}^{2}-d_{{\rm t}n}^{2})-{\rm Im}\left(\frac{2}{D_{n}^{3}}\right)d_{{\rm s}n}d_{{\rm t}n}\right]\phi,
\end{align}
which can be applied in weak magnetic fields, since we have neglected orbital effects.
There exist the {\it paramagnetic} and the interference contributions to $n_{\rm s}$ other than the ordinary diamagnetic one, since the signs of $d_{\rm s}^{2}$ and $d_{\rm t}^{2}$ terms are opposite.

In what follows, we adopt relatively large $\lambda$ in order to emphasize characteristic features of the even-odd-frequency mixing.
However, qualitative features of the results are not altered for weaker $\lambda$ and $h$.
We use the unit of energy as $\omega_{\rm E}=1$.

\subsection{Single-gap approximation in the strong-coupling limit}

Let us first consider the strong-coupling limit, $\lambda\to\infty$ in order to grasp an outline of the solutions.
In this limit, $n=0$ and $-1$ components, i.e., $\Delta_{0}=\Delta_{-1}^{*}\equiv \Delta_{\rm s}+i\Delta_{\rm t}$ dominate over other components\cite{Carbotte90,Cappelluti07}.
As will be shown below, this simple approximation provides important information on the structure of the free-energy functional and qualitatively similar behaviors to those obtained by solving the full Eliashberg equations\cite{Matsumoto12}.
The approximate free-energy difference is given by
\begin{equation}
\Delta\Omega[\Delta_{\rm s},\Delta_{\rm t}]=
\frac{2\rho_{\rm F}}{\lambda}\phi\left(\frac{\Delta_{\rm s}^{2}}{u_{\rm s}}-\frac{\Delta_{\rm t}^{2}}{u_{\rm t}}\right)
-4\rho_{\rm F}\omega_{0}\,{\rm Re}[D_{0}-Z_{0}\omega_{0}].
\end{equation}
Here, the self energy is evaluated in the normal state for simplicity as $Z_{0}=1+\lambda$.
$D_{0}=\sqrt{(Z_{0}\omega_{0}-ih)^{2}+\phi(\Delta_{\rm s}+i\Delta_{\rm t})^{2}}$.
The normalized attractions in the singlet and the triplet channels are introduced as
\begin{align}
&u_{\rm s}(t)\equiv v_{0,0}+v_{0,-1}=\frac{2+t^{2}}{1+t^{2}},
\\
&u_{\rm t}(t)\equiv v_{0,0}-v_{0,-1}=\frac{t^{2}}{1+t^{2}},
\end{align}
with $t\equiv 2\pi T/\omega_{\rm E}$.
The corresponding gap equations and the equilibrium free-energy difference are given by
\begin{align}
&\Delta_{\rm s}=\lambda u_{\rm s}\omega_{0}\,{\rm Re}\left[\frac{\Delta_{\rm s}+i\Delta_{\rm t}}{D_{0}}\right],
\notag\\
&\Delta_{\rm t}=\lambda u_{\rm t}\omega_{0}\,{\rm Im}\left[\frac{\Delta_{\rm s}+i\Delta_{\rm t}}{D_{0}}\right],
\label{gapeq}
\\
&\Delta \Omega_{\rm s}=
2\rho_{\rm F}\omega_{0}\,{\rm Re}\biggl[\frac{\phi (\Delta_{\rm s}+i\Delta_{\rm t})^{2}-\omega_{0}(1-Z_{0})(Z_{0}\omega_{0}-ih)}{D_{0}}
\notag\\&\quad\quad\quad\quad\quad\quad\quad\quad\quad\quad
+\omega_{0}(1-Z_{0})-2(D_{0}-Z_{0}\omega_{0})\biggr].
\end{align}

\begin{figure}[thb]
\begin{center}
\includegraphics[width=8.5cm]{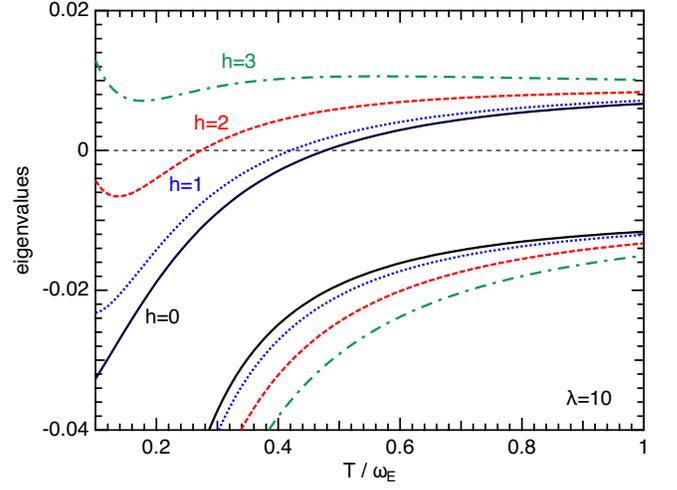}
\end{center}
\caption{(Color online) The $T$ dependence of eigenvalues of ${\cal K}(T,h)$. One eigenvalue is always negative, while the other changes its sign at $T_{c}(h)$ except for $h/\omega_{\rm E}=3$.}
\label{g_evf}
\end{figure}

Expanding $\Omega$ with respect to $\Delta_{\rm s}$ and $\Delta_{\rm t}$, we obtain the lowest-order Landau free energy as
\begin{gather}
\Delta\Omega[\Delta_{\rm s},\Delta_{\rm t}]=
2\rho_{\rm F}\phi\,
(\Delta_{\rm s},\,\,\Delta_{\rm t})
{\cal K}
\begin{pmatrix} \Delta_{\rm s} \\ \Delta_{\rm t} \end{pmatrix}+{\cal O}(\Delta^{4}),
\\
{\cal K}(T,h)=
\begin{pmatrix}
1/\lambda u_{\rm s}-A & B \\
B & -\left(1/\lambda u_{\rm t}-A\right)
\end{pmatrix},
\label{fekernel}
\end{gather}
with
\begin{equation}
A=\frac{Z_{0}\omega_{0}^{2}}{(Z_{0}\omega_{0})^{2}+h^{2}},
\quad
B=\frac{\omega_{0}h}{(Z_{0}\omega_{0})^{2}+h^{2}}.
\end{equation}
Note that the minus sign appears in the (2,2) component of (\ref{fekernel}) due to the relation, (\ref{ddpcond}).
The superconducting instability at $T_{c}(h)$ is determined by the condition $|{\cal K}(T_{c},h)|=0$, or equivalently, by the condition such that one of the eigenvalues becomes zero.

Figure~\ref{g_evf} shows the $T$ dependence of eigenvalues of ${\cal K}(T,h)$ at $h/\omega_{\rm E}=0$, $1$, $2$, and $3$ for $\lambda=10$.
One eigenvalue coming mainly from the triplet channel is always negative as a consequence of (\ref{ddpcond}), while the other changes its sign at $T_{c}(h)$.
This behavior indicates that the normal state is not characterized by the minimum as usual, but by the saddle point of the {\it constrained} free-energy functional above $T_{c}$.
Then, the saddle point at the origin turns into the unstable maximum (stable minimum) below $T_{c}$ in the case of $\phi=+1$ ($\phi=-1$).
Therefore, the choice of $\phi=+1$ describes a realistic phase transition to the superconducting state.
In other words, $\phi=+1$ is chosen since the singlet pairing is the {\it majority} component in the present case (the pure triplet pairing is suppressed completely by the mass-enhancement effect).
In the absence of the magnetic field where the singlet and the triplet channels are decoupled, the sign $\phi$ can be chosen independently in the singlet ($\phi=+1$) and the triplet ($\phi=-1$) channels.
However, the finite coupling in the presence of the magnetic field requires the opposite sign of $\phi$ between two components, yielding the energy loss in the triplet channel, even though the interaction is originally attractive in the triplet channel.
Although the relative-sign requirement (\ref{ddpcond}) leads to higher free energy than the case of neglecting the odd-frequency triplet component as in the past studies, $\Delta_{\rm t}=0$ line in $\Delta_{\rm s}$-$\Delta_{\rm t}$ plane is no longer a stationary solution of the free energy in the presence of magnetic fields, and the saddle point of the constrained free-energy functional gives a stable solution.
The upturn of the eigenvalue in lower temperatures suggests a reentrant behavior of the superconducting phase.
For higher fields (e.g. $h=3$), no sign changes occur and the system remains the normal state.

\begin{figure}[htb]
\begin{center}
\includegraphics[width=8.5cm]{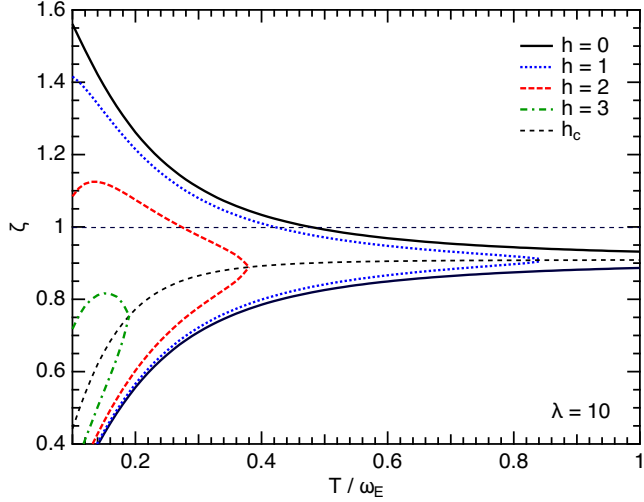}
\end{center}
\caption{(Color online) The $T$ dependence of two real eigenvalues $\zeta$ of the kernel in the linearized gap equation. The dashed line with the label $h_{c}$ represents $\zeta$ for $h_{\rm c}(T)$ where two eigenvalues coincide. $\zeta(T_{c})=1$ gives the transition temperature.}
\label{g_lam0t}
\end{figure}

\begin{figure}[htb]
\begin{center}
\includegraphics[width=8.5cm]{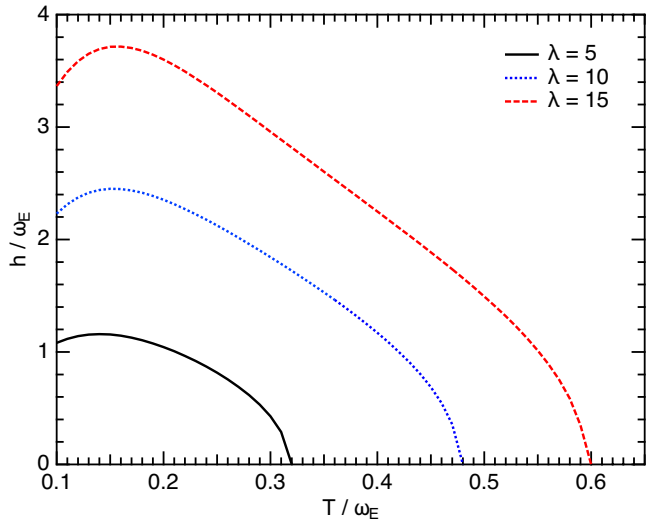}
\end{center}
\caption{(Color online) The $T$-$h$ phase diagram for several coupling constants. The phase boundary shows reentrant behavior in high fields.}
\label{g_htphase}
\end{figure}

\begin{figure*}[tbh]
\begin{center}
\includegraphics[width=15cm]{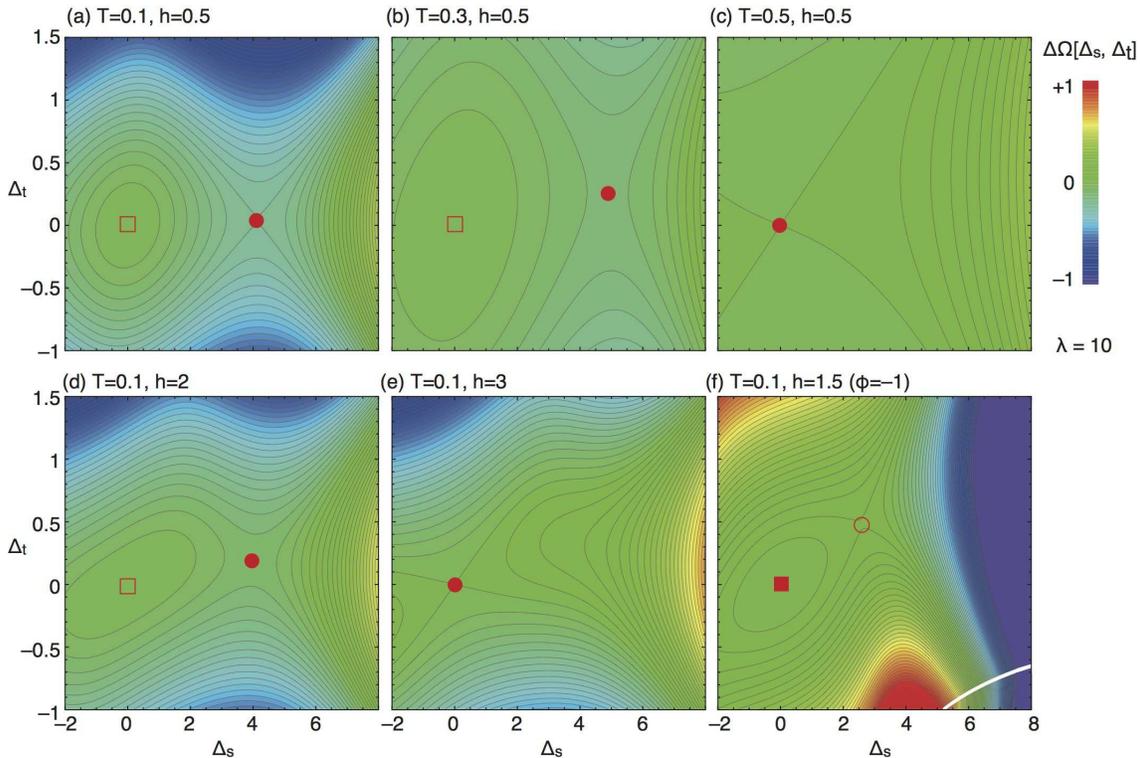}
\end{center}
\caption{(Color online) The landscape of the constrained free-energy functional in unit of $2\rho_{\rm F}$ measured from that of the normal state. (a)-(c) the $T$ dependence at $h/\omega_{\rm E}=0.5$, (d), (e) the $h$ dependence at $T/\omega_{\rm E}=0.1$, (f) the case of $\phi=-1$ at $T/\omega_{\rm E}=0.1$, $h/\omega_{\rm E}=1.5$. The closed circles represent the saddle points corresponding to the solutions of the gap equation (see, Fig.~\ref{g_gap}). The open circle and squares indicate other saddle point and maxima, whose free energies are higher than the saddle points indicated by the closed circles. The closed square represents the minimum of $\Delta\Omega$ in the case of $\phi=-1$. On the white line in the right-lower region in (f), ${\rm Re}(D_{0})=0$ at which the derivative is discontinuous.}
\label{fmapl}
\end{figure*}

\begin{figure}[thb]
\begin{center}
\includegraphics[width=8.5cm]{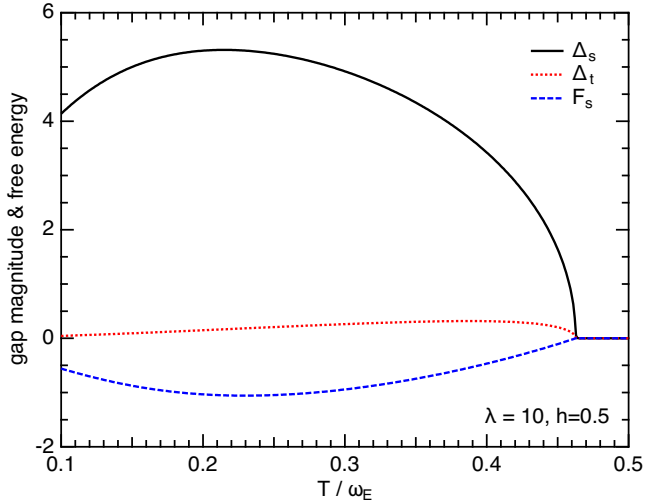}
\end{center}
\caption{(Color online) The $T$ dependence of the gap functions and the equilibrium free energy in unit of $2\rho_{\rm F}$ at $h/\omega_{\rm E}=0.5$.}
\label{g_gap}
\end{figure}

\begin{figure}[tht]
\begin{center}
\includegraphics[width=8.5cm]{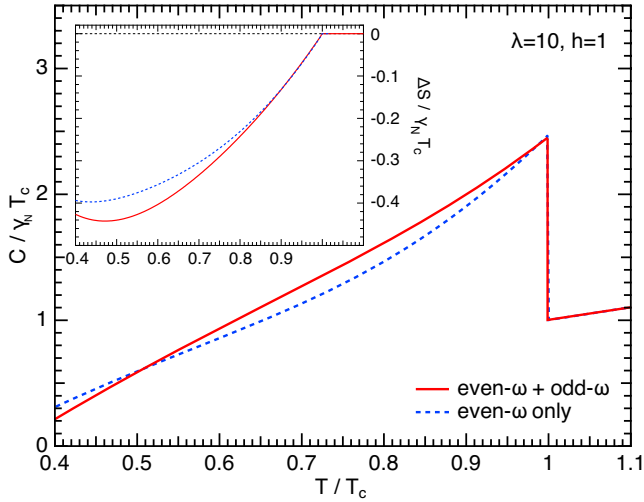}
\end{center}
\caption{(Color online) The comparison of the $T$-dependences of the specific heat with or without the odd-frequency component. The entropy difference is shown in the inset.}
\label{g_ct}
\end{figure}

\begin{figure}[thb]
\begin{center}
\includegraphics[width=8.5cm]{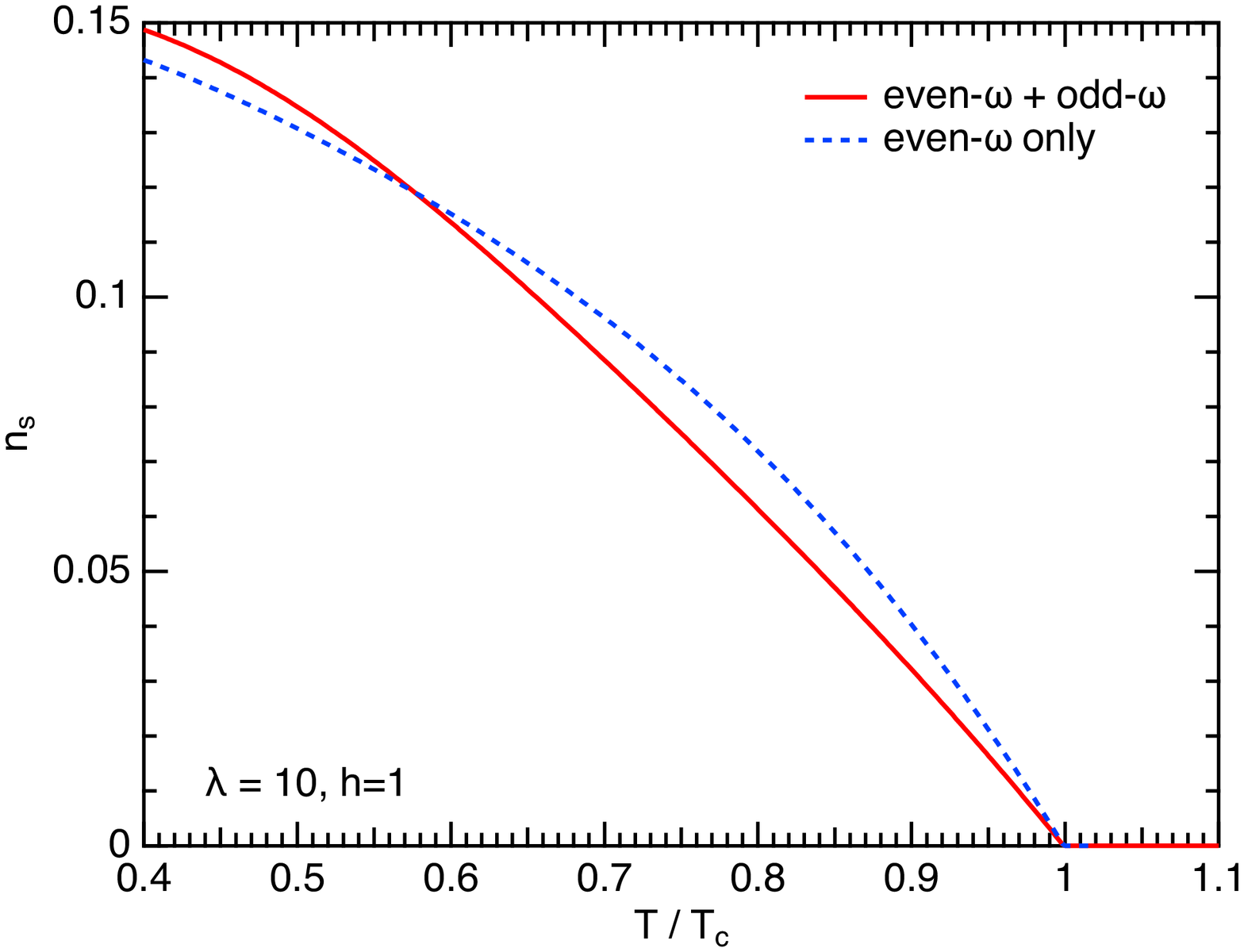}
\end{center}
\caption{(Color online) The comparison of the $T$-dependence of the superfluid density with or without the odd-frequency component.}
\label{g_ns}
\end{figure}

The superconducting instability can also be observed by the linearized gap equation,
\begin{equation}
\zeta
\begin{pmatrix} \Delta_{\rm s} \\ \Delta_{\rm t} \end{pmatrix}
=\lambda
\begin{pmatrix}
u_{\rm s}A & -u_{\rm s}B \\
u_{\rm t}B & u_{\rm t}A
\end{pmatrix}
\begin{pmatrix} \Delta_{\rm s} \\ \Delta_{\rm t} \end{pmatrix},
\end{equation}
where the eigenvalue $\zeta=1$ at $T_{c}$ signals the superconducting instability.
The eigenvalues are given by
\begin{equation}
\zeta=\lambda A\left[1\pm\frac{1}{1+t^{2}}\sqrt{1-\frac{t^{2}(2+t^{2})h^{2}}{(Z_{0}\omega_{0})^{2}}}\right],
\end{equation}
which are shown in Fig.~\ref{g_lam0t} for $h/\omega_{\rm E}=0$, $1$, $2$ and $3$ at $\lambda=10$.
The dashed line with the label $h_{c}$ represents $\zeta$ for $h_{\rm c}(T)$ where two eigenvalues coincide.
In the presence of magnetic fields, the kernel of the linearized gap equation becomes non-hermite, leading to the complex eigenvalues when
\begin{equation}
h>h_{\rm c}\equiv \frac{Z_{0}\omega_{0}}{t\sqrt{2+t^{2}}}.
\end{equation}
The eigenvalues shown in Fig.~\ref{g_lam0t} provide qualitatively similar behaviors to those obtained by solving the full linearized Eliashberg equations, (\ref{eeqs})\cite{Matsumoto12}.

The $T$-$h$ phase diagram determined by $|{\cal K}(T_{c},h)|=0$ or $\zeta(T_{c},h)=1$ for $\lambda=5$, $10$, and $15$ is shown in Fig.~\ref{g_htphase}.
As was expected, the superconducting phase boundary shows the reentrant behavior in high fields.
$T_{c}(h)$ is considerably suppressed as compared with the case of neglecting the odd-frequency triplet component\cite{Matsumoto12}.

Figure~\ref{fmapl} shows the typical landscapes of the constrained free-energy functional $\Delta\Omega[\Delta_{\rm s},\Delta_{\rm t}]/2\rho_{\rm F}$ for $\lambda=10$.
The upper panels (a)-(c) show the $T$ dependence at $h/\omega_{\rm E}=0.5$, and the lower panels (d), (e) show the $h$ dependence at $T/\omega_{\rm E}=0.1$.
Figure~\ref{fmapl}(f) is the case of $\phi=-1$ at $T/\omega_{\rm E}=0.1$ and $h/\omega_{\rm E}=1.5$.
The closed circles represent the saddle points corresponding to the solutions of the gap equation (see, Fig.~\ref{g_gap}). The open circle and squares indicate other saddle point and maxima, whose free energies are higher than the saddle points represented by the closed circles.
The closed square represents the minimum of $\Delta\Omega$ in the case of $\phi=-1$.
By the comparison between Figs.~\ref{fmapl} and \ref{g_gap}, we can see that both the normal and the superconducting states are characterized by the {\it saddle points} of the {\it constrained} free-energy functional with the choice of $\phi=+1$.
Note that an arbitrary initial condition always converges to the solution of the Eliashberg equations that corresponds to the saddle point of the constrained free-energy functional in the case of $\phi=+1$.
On the other hand, in the case of $\phi=-1$, the energy of the saddle point is always higher than that of the origin in the constrained free-energy functional, which corresponds to the normal state.
This is the consequence of the wrong choice of $\phi$.

\subsection{Comparison with or without the triplet component}

Let us now discuss some physical quantities by solving the full Eliashberg equations, (\ref{eeqs}), and compare with those obtained by neglecting the odd-frequency component (${\rm Im}\,\Delta_{n}=d_{tn}$ is forced to be zero during the computation), which corresponds to the limit of vanishing attraction in the triplet channels.
Figure~\ref{g_ct} shows the comparison of the specific heat $C(T)$ and the entropy difference $\Delta S(T)$ from the normal state.
Here, $\gamma_{\rm N}=2\rho_{\rm F}\pi^{2}(1+\lambda)/3$ is the renormalized Sommerfeld coefficient in the normal state.
Although $T_{c}(h)$ is considerably different from each other, the discontinuities at $T_{c}$ show almost no difference ($\Delta C/\gamma_{\rm N}T_{c}\sim1$) and the slopes of $C(T)$ below $T_{c}$ differ slightly from each other.
Namely, the entropy release in the mixed-frequency solution is more gradual than that in the pure singlet solution.
Note that the correct mixed-frequency solution has lower entropy than that of the pure singlet solution.

The comparison of the superfluid density $n_{s}(T)$ is shown in Fig.~\ref{g_ns}.
The evolutions of $n_{s}(T)$ near $T_{c}$ are linear in both cases, but the slope of the mixed-frequency solution is slightly smaller than that of the pure singlet solution.
In low enough temperatures, however, the superfluid density of the mixed-frequency solution is larger than that of the pure singlet solution.
This behavior can be interpreted by a paramagnetic contribution from the induced odd-frequency component in the vicinity of $T_{c}$, which suppresses the Meissner screening slightly.
The majority singlet component gives the ordinary diamagnetic Meissner current, which eventually dominates over the paramagnetic contribution from the induced odd-frequency component at low temperatures.
The existence of the paramagnetic Meissner effect is extensively discussed in the spatially nonuniform systems at the surface/interface\cite{Bergeret05,Tanaka12}.

\section{Summary}

We have investigated general structure of the free-energy functional for the coexistence state of the singlet and triplet pairings with the even-odd-frequency mixing.
The real condition of the free-energy functional for the coexistence state requires the relation between the gap functions and their particle-hole converted counterparts as,
\[
d_{0}^{+}(k)=\phi d_{0}(k)^{*},
\quad
{\bm d}^{+}(k)=-\phi{\bm d}(k)^{*},
\]
where $\phi$ takes either $+1$ or $-1$ depending on which component is majority, i.e., $\phi=+1$ for the singlet majority and $\phi=-1$ for the triplet majority, in order to lower the resultant equilibrium free energy via the interference effect.
In the pure singlet or the triplet state, this requirement reduces to the ordinary relations\cite{Kusunose11a}, $d_{0}^{+}(k)=d_{0}(k)^{*}$ or ${\bm d}^{+}(k)={\bm d}(k)^{*}$.

The opposite signs in the singlet and the triplet channels lead to the anomalous structure of the free-energy functional, which is constrained by the above relations, and the equilibrium normal and superconducting states are characterized not by the minimum, but by the saddle point of the constrained free-energy functional.
The minority component with the extraordinary relation, $d_{\alpha}^{+}(k)=-d_{\alpha}(k)^{*}$ suppresses the superconducting state yielding lower $T_{c}$ as compared with those obtained by neglecting the minority component.
It also reflects in property of various physical quantities such as the specific heat and the superfluid density.

Using the simplest example of the ordinary $s$-wave singlet state mixed with the triplet state with the odd-frequency dependence under magnetic fields, we have demonstrated the saddle-point structure of the constrained free-energy functional, and the induced odd-frequency triplet component (the minority component with the opposite sign) alters the $T$ dependences of the specific heat and the superfluid density as compared with those obtained by neglecting the odd-frequency triplet component.
It would be interesting to detect such effect under the broken time-reversal symmetry in strong-coupling superconductors.

\section*{Acknowledgments}

We acknowledge valuable discussions with Y. Fuseya, K. Miyake, and Y. Tanaka.
This work is supported by a Grant-in-Aid for Scientific Research C (No. 23540414) from the Japan Society for the Promotion of Science.
One of the authors (H.K.) is supported by a Grant-in-Aid for Scientific Research on Innovative Areas ``Heavy Electrons" (No. 20102008) of The Ministry of Education, Culture, Sports, Science, and Technology (MEXT), Japan.

\appendix
\section{Derivation of the free-energy functional, and the Green's functions}
\subsection{The free-energy functional}
Here, we show the derivation of the free-energy functional in terms of the spin-decomposed components.
Let us start from the expression (\ref{glf}).
For $\Theta[\hat{\Sigma}]$, it is straightforward to obtain,
\begin{align}
\frac{\Theta}{2}&=-\frac{1}{4}\sum_{kk'}W(k-k'){\rm Tr}\left[\hat{\Sigma}(k)\hat{\rho}_{3}\hat{\Sigma}(k')\hat{\rho}_{3}\right]
\notag\\&
=-\frac{1}{4}\sum_{kk'}W(k-k'){\rm Tr}\left[
\begin{pmatrix}
\Sigma(k) & \Delta(k) \\ \Delta^{+}(k) & \Sigma^{+}(k)
\end{pmatrix}
\begin{pmatrix}
\Sigma(k') & -\Delta(k') \\ -\Delta^{+}(k') & \Sigma^{+}(k')
\end{pmatrix}
\right]
\notag\\&
=-\frac{1}{2}\sum_{kk'}W(k-k'){\rm tr}\left[
\Sigma(k)\Sigma(k')-\Delta(k)\Delta^{+}(k')
\right]
\notag\\&
=\sum_{kk'}W(k-k')\left[d_{0}(k)d_{0}^{+}(k')+{\bm d}(k)\cdot{\bm d}^{+}(k')
\right.\notag\\&\quad\quad\quad\quad\quad\quad\quad\quad\left.
-\Sigma_{0}(k)\Sigma_{0}(k')-{\bm\Sigma}(k)\cdot{\bm\Sigma}(k')
\right].
\end{align}

In the first term of (\ref{glf}), we use the identity, ${\rm Tr}\ln(\cdots)=\ln{\rm det}(\cdots)$.
For a general $4\times4$ matrix that consists of $2\times2$ matrices, $A,\cdots, D$, can be decomposed as
\begin{gather}
\begin{pmatrix}
A & B \\ C & D
\end{pmatrix}
=
\begin{pmatrix}
1 & BD^{-1} \\ 0 & 1
\end{pmatrix}
\begin{pmatrix}
A(1-A^{-1}BD^{-1}C) & 0 \\ 0 & D
\end{pmatrix}
\begin{pmatrix}
1 & 0 \\ D^{-1}C & 1
\end{pmatrix}.
\label{matid}
\end{gather}
Using this identity, and omitting $k$ for notational simplicity, we obtain
\begin{align}
X(k)\equiv{\rm det}&
\begin{pmatrix}
-z & \Delta \\ \Delta^{+} & -z^{+}
\end{pmatrix}
=ww^{+}\biggl|1-(z)^{-1}\Delta (z^{+})^{-1}\Delta^{+}\biggr|,
\label{xex}
\end{align}
where
\begin{equation}
w=|z|=z_{0}^{2}-{\bm z}^{2},
\quad
w^{+}=|z^{+}|=z_{0}^{+2}-{\bm z}^{+2}.
\end{equation}
By the straightforward calculation, we have
\begin{align}
&(z)^{-1}\Delta=\frac{1}{w}\left[
\alpha_{0}\tau_{0}+{\bm\alpha}\cdot{\bm\tau}
\right],
\\
&(z^{+})^{-1}\Delta^{+}=\frac{1}{w^{+}}\left[
\alpha_{0}^{+}\tau_{0}^{\dagger}+{\bm\alpha}^{+}\cdot{\bm\tau}^{\dagger}
\right],
\end{align}
where we have defined
\begin{align}
&\alpha_{0}=z_{0}d_{0}-{\bm z}\cdot{\bm d},
\notag\\
&{\bm\alpha}=z_{0}{\bm d}-{\bm z}d_{0}-i({\bm z}\times{\bm d}),
\notag\\
&\alpha_{0}^{+}=z_{0}^{+}d_{0}^{+}+{\bm z}^{+}\cdot{\bm d}^{+},
\notag\\
&{\bm\alpha}^{+}=z_{0}^{+}{\bm d}^{+}+{\bm z}^{+}d_{0}^{+}+i({\bm z}^{+}\times{\bm d}^{+}).
\label{aldef}
\end{align}
Using these expressions, we obtain
\begin{align}
P\equiv 1-(z)^{-1}\Delta (z^{+})^{-1}\Delta^{+}=\frac{D_{0}\sigma_{0}-({\bm D}+{\bm M})\cdot{\bm\sigma}}{ww^{+}},
\label{deter}
\end{align}
where
\begin{align}
&D_{0}=ww^{+}-(\alpha_{0}\alpha_{0}^{+}+{\bm\alpha}\cdot{\bm\alpha}^{+}),
\notag\\
&{\bm D}=\alpha_{0}{\bm\alpha}^{+}+{\bm\alpha}\alpha_{0}^{+},
\quad
{\bm M}=i({\bm\alpha}\times{\bm\alpha}^{+}).
\end{align}
Putting these expressions into (\ref{xex}) and using ${\bm D}\cdot{\bm M}=0$, we have
\begin{equation}
X(k)=\frac{D_{0}^{2}-{\bm D}^{2}-{\bm M}^{2}}{ww^{+}}.
\label{exdet}
\end{equation}
The final expression of the free-energy functional is given by
\begin{multline}
\Omega[\hat{\Sigma}]=\sum_{kk'}W(k-k')\left[d_{0}(k)d_{0}^{+}(k')+{\bm d}(k)\cdot{\bm d}^{+}(k')
\right.\\\left.
-\Sigma_{0}(k)\Sigma_{0}(k')-{\bm\Sigma}(k)\cdot{\bm\Sigma}(k')
\right]-\frac{T}{2}\sum_{k}\ln\left(\frac{X(k)}{X_{0}(k)}\right),
\label{expfef}
\end{multline}
where $X_{0}(k)$ is the non-interacting value of $X(k)$.

The stationary condition of $\Omega[\hat{\Sigma}]$ gives the Eliashberg equations,
\begin{align}
&\Sigma_{\alpha}(k)=\frac{T}{2}\sum_{k'}V(k-k')g_{\alpha}(k'),
\quad
(\alpha=0,1,2,3),
\label{ee1}
\\
&d_{\alpha}(k)=\frac{T}{2}\sum_{k'}V(k-k')f_{\alpha}(k'),
\notag\\
&d_{\alpha}^{+}(k)=\frac{T}{2}\sum_{k'}V(k-k')f_{\alpha}^{+}(k'),
\label{ee2}
\end{align}
where we have defined the Green's functions as
\begin{align}
&g_{\alpha}(k)=-\frac{\delta\ln X(k)}{\delta\Sigma_{\alpha}(k)},
\\
&f_{\alpha}(k)=\frac{\delta\ln X(k)}{\delta d_{\alpha}^{+}(k)},
\quad
f_{\alpha}^{+}(k)=\frac{\delta\ln X(k)}{\delta d_{\alpha}(k)}.
\end{align}
Note that $\Sigma(k)$ and $\Sigma^{+}(k)$ in $X(k)$ should be treated as independent quantities in the functional derivative.
Their explicit expressions are given later.

Using the solution of the Eliashberg equations, we obtain the equilibrium value as
\begin{align}
\frac{\Theta_{\rm s}}{2}&
=\frac{T}{2}\sum_{k}\left[f_{0}(k)d_{0}^{+}(k)+{\bm f}(k)\cdot{\bm d}^{+}(k)
\right.\notag\\&\quad\quad\quad\quad\quad\quad\quad\quad\left.
-g_{0}(k)\Sigma_{0}(k)-{\bm g}(k)\cdot{\bm\Sigma}(k)
\right],
\end{align}
and the explicit equilibrium free energy is given by (\ref{frees}) with use of the relation, (\ref{ddpcond}).

As we discussed in detail in \S II.B, the real condition of the free-energy functional requires the relation, (\ref{ddpcond}),
\[
d_{0}^{+}(k)=\phi d_{0}(k)^{*},
\quad
{\bm d}^{+}(k)=-\phi{\bm d}(k)^{*},
\quad
(\phi=+1,\text{ or }-1).
\]
With these constraints, we have
\begin{equation}
\alpha_{0}^{+}=-\phi\alpha_{0}^{*},
\quad
{\bm\alpha}^{+}=\phi{\bm\alpha}^{*}.
\end{equation}
From (\ref{ee2}) and (\ref{ddpcond}), it can be shown the relation, (\ref{frel}),
\[
f_{0}^{+}(k)=\phi f_{0}(k)^{*},
\quad
{\bm f}^{+}(k)=-\phi {\bm f}(k)^{*}.
\]

\subsection{The lowest-order Landau expansion}
In order to obtain the lowest-order expression of the Landau expansion, we expand the last term of (\ref{expfef}) with respect to $\alpha_{\alpha}$ and $\alpha^{+}_{\alpha}$ ($\alpha=0,1,2,3$), then we have
\begin{align}
-\frac{T}{2}\ln\left(\frac{X(k)}{X_{0}(k)}\right)&-c_{\rm N}
\sim
-T\ln\left(\frac{D_{0}}{ww^{+}}\right)
\notag\\&
=
-T\ln\left(1-\frac{\alpha_{0}\alpha_{0}^{+}+{\bm\alpha}\cdot{\bm\alpha}^{+}}{ww^{+}}\right)
\notag\\&
\sim
\frac{T}{ww^{+}}\left(\alpha_{0}\alpha_{0}^{+}+{\bm\alpha}\cdot{\bm\alpha}^{+}\right),
\end{align}
where $c_{\rm N}=-(T/2)\sum_{k}\ln(ww^{+}/X_{0})$.
Using the explicit expression,
\begin{align}
\alpha_{0}\alpha_{0}^{+}&+{\bm\alpha}\cdot{\bm\alpha}^{+}
=
(z_{0}z_{0}^{+}-{\bm z}\cdot{\bm z}^{+})(d_{0}d_{0}^{+}+{\bm d}\cdot{\bm d}^{+})
\notag\\&\quad
+({\bm z}\times{\bm d})\cdot({\bm z}^{+}\times{\bm d}^{+})+({\bm z}\times{\bm d}^{+})\cdot({\bm z}^{+}\times{\bm d})
\notag\\&\quad
-{\bm n}_{0}\cdot {\bm m}_{0}
+{\bm n}_{1}\cdot{\bm m}_{1}
+{\bm n}_{2}\cdot{\bm m}_{2},
\notag\\&
=(z_{0}z_{0}^{+}-{\bm z}\cdot{\bm z}^{+})d_{0}d_{0}^{+}
+(z_{0}z_{0}^{+}+{\bm z}\cdot{\bm z}^{+}){\bm d}\cdot{\bm d}^{+}
\notag\\&\quad
-\frac{1}{2}\sum_{i,j}^{1,2,3}N_{ij}M_{ij}
-{\bm n}_{0}\cdot {\bm m}_{0}
+{\bm n}_{1}\cdot{\bm m}_{1}
+{\bm n}_{2}\cdot{\bm m}_{2},
\label{aaaa}
\end{align}
where ${\bm n}_{i}$ and ${\bm m}_{i}$ are defined by (\ref{nvec}) and (\ref{mvec}), we obtain the lowest-order Landau expansion as (\ref{glex2}).
The symmetric matrices are defined by $N_{ij}=z_{i}z_{j}^{+}+z_{i}^{+}z_{j}$ and $M_{ij}=d_{i}d^{+}_{j}+d_{i}^{+}d_{j}$, respectively.

\subsection{The Green's functions}

The stationary condition of the free-energy functional leads to the Dyson (Eliashberg) equation,
\begin{equation}
\begin{pmatrix}
G(k) & -F(k) \\
-F^{+}(k) & G^{+}(k)
\end{pmatrix}=
\begin{pmatrix}
z(k) & -\Delta(k) \\
-\Delta^{+}(k) & z^{+}(k)
\end{pmatrix}^{-1}.
\end{equation}
Thus, by inverting the matrix explicitly, we obtain the expressions of the Green's function.
For this purpose, we again use the identity, (\ref{matid}), then we express the inverse of a matrix as
\begin{align}
\begin{pmatrix}
A & B \\ C & D
\end{pmatrix}^{-1}
&=
\begin{pmatrix}
1 & 0 \\ D^{-1}C & 1
\end{pmatrix}^{-1}
\begin{pmatrix}
A-BD^{-1}C & 0 \\ 0 & D
\end{pmatrix}^{-1}
\begin{pmatrix}
1 & BD^{-1} \\ 0 & 1
\end{pmatrix}^{-1}
\notag\\&
=
\begin{pmatrix}
1 & 0 \\ -D^{-1}C & 1
\end{pmatrix}
\begin{pmatrix}
(A-BD^{-1}C)^{-1} & 0 \\ 0 & D^{-1}
\end{pmatrix}
\begin{pmatrix}
1 & -BD^{-1} \\ 0 & 1
\end{pmatrix}
\notag\\&
=
\begin{pmatrix}
(A-BD^{-1}C)^{-1} & (C-DB^{-1}A)^{-1} \\ (B-AC^{-1}D)^{-1} & (D-CA^{-1}B)^{-1}
\end{pmatrix}.
\end{align}
Using this formula, we have
\begin{align}
&
G(k)=\left[1-(z)^{-1}\Delta(z^{+})^{-1}\Delta^{+}\right]^{-1}(z)^{-1}=(zP)^{-1},
\\
&
F(k)=-\left[1-(z)^{-1}\Delta(z^{+})^{-1}\Delta^{+}\right]^{-1}(z)^{-1}\Delta(z^{+})^{-1}=-G\Delta(z^{+})^{-1}.
\end{align}
Using (\ref{deter}) and after some manipulations, the explicit expressions are given by
\begin{align}
&g_{0}(k)=\frac{1}{X}\left[w^{+}z_{0}-(d_{0}d_{0}^{+}+{\bm d}\cdot{\bm d}^{+})z_{0}^{+}+({\bm m}_{0}-{\bm m}_{2})\cdot{\bm z}^{+}\right],
\notag\\
&{\bm g}(k)=\frac{1}{X}\biggl[-w^{+}{\bm z}
+(d_{0}d_{0}^{+}-{\bm d}\cdot{\bm d}^{+}){\bm z}^{+}
+({\bm m}_{0}+{\bm m}_{2})z_{0}^{+}
\notag\\&\quad\quad\quad
+({\bm z}^{+}\cdot{\bm d}){\bm d}^{+}
+({\bm z}^{+}\cdot{\bm d}^{+}){\bm d}
+i({\bm m}_{1}\times{\bm z}^{+})\biggr],
\label{gf1}
\\
&f_{0}(k)=\frac{1}{X}\biggl[
-(d_{0}^{2}-{\bm d}^{2})d_{0}^{+}
+(z_{0}z_{0}^{+}-{\bm z}\cdot{\bm z}^{+})d_{0}
+({\bm n}_{1}+{\bm n}_{2})\cdot{\bm d}\biggr],
\notag\\
&{\bm f}(k)=\frac{1}{X}\biggl[
(d_{0}^{2}-{\bm d}^{2}){\bm d}^{+}
+(z_{0}z_{0}^{+}+{\bm z}\cdot{\bm z}^{+}){\bm d}
-({\bm n}_{1}-{\bm n}_{2})d_{0}
\notag\\&\quad\quad\quad
-({\bm z}^{+}\cdot{\bm d}){\bm z}
-({\bm z}\cdot{\bm d}){\bm z}^{+}
-i({\bm n}_{0}\times{\bm d})\biggr].
\label{gf2}
\end{align}

\section{Effect of the Rashba spin-orbit coupling}

Here, we consider briefly the effect of the Rashba antisymmetric spin-orbit coupling\cite{Bauer12}, ${\bm\gamma}_{\bm k}=-{\bm\gamma}_{-{\bm k}}=\lambda_{\rm R}({\bm k}\times\hat{\bm z})=\lambda_{\rm R}(k_{y},-k_{x},0)$.
In this case, we just replace the spin part with
\begin{align}
{\bm z}(k)=-{\bm\gamma}_{\bm k}-{\bm\Sigma}(k).
\end{align}
In the presence of the Rashba spin-orbit coupling, ${\bm z}^{+}(k)=-{\bm z}(k)^{*}$ does not hold in general, but $ww^{+}=|w|^{2}$ still holds.

The symmetry properties of the symmetric matrix $N_{ij}$ and the symmetry-lowering vectors ${\bm n}_{i}$ in (\ref{aaaa}) are extracted as
\begin{align}
&
N(k)\sim\lambda_{\rm R}^{2}\begin{pmatrix}
k_{y}^{2} & k_{x}k_{y} & 0 \\
k_{x}k_{y} & k_{x}^{2} & 0 \\
0 & 0 & 0
\end{pmatrix},
\\
&{\bm n}_{0}(k)\sim i\omega_{n}{\bm\gamma}_{\bm k}=i\omega_{n}\lambda_{\rm R}(k_{y},-k_{x},0),
\\
&{\bm n}_{1}(k)\sim 0,
\\
&{\bm n}_{2}(k)\sim \xi_{\bm k}{\bm\gamma}_{\bm k}=\xi_{\bm k}\lambda_{\rm R}(k_{y},-k_{x},0).
\end{align}
In order to gain the interference energies, and the free energy is to be real, we should have the relation,
\begin{equation}
d_{0,1,2}^{+}(k)=\phi d_{0,1,2}(k)^{*},
\quad
d_{3}^{+}(k)=-\phi d_{3}(k)^{*}.
\end{equation}
Then, the conjugate quantities become
\begin{align}
&
M(k)=2\phi\begin{pmatrix}
|d_{1}|^{2} & {\rm Re}(d_{1}d_{2}^{*}) & -i\,{\rm Im}(d_{1}d_{3}^{*}) \\
{\rm Re}(d_{1}d_{2}^{*}) & |d_{2}|^{2} & -i\,{\rm Im}(d_{2}d_{3}^{*}) \\
-i\,{\rm Im}(d_{1}d_{3}^{*}) & -i\,{\rm Im}(d_{2}d_{3}^{*}) & -|d_{3}|^{2}
\end{pmatrix},
\\
&{\bm m}_{0}(k)=-2\phi[i\,{\rm Re}(d_{2}d_{3}^{*}),-i\,{\rm Re}(d_{3}d_{1}^{*}),{\rm Im}(d_{1}d_{2}^{*})],
\\
&{\bm m}_{1}(k)=-2\phi[i\,{\rm Im}(d_{0}d_{1}^{*}),i\,{\rm Im}(d_{0}d_{2}^{*}),-{\rm Re}(d_{0}d_{3}^{*})],
\\
&{\bm m}_{2}(k)=2\phi[{\rm Re}(d_{0}d_{1}^{*}),{\rm Re}(d_{0}d_{2}^{*}),-i\,{\rm Im}(d_{0}d_{3}^{*})].
\end{align}
The relative phases among $d_{\alpha}$ should be taken as real, and ${\rm Im}(d_{\alpha}d_{\beta}^{*})$ vanishes.
Note that the frequency mixing occurs between $(d_{0},d_{1},d_{2})$ and $d_{3}$ components via the ${\bm n}_{0}$ term, while the spatial parity mixing occurs between $(d_{0},d_{3})$ and $(d_{1},d_{2})$ components via the $N_{ij}$, ${\bm n}_{0}$ and ${\bm n}_{2}$ terms.
Consequently, the spin mixing occurs between $d_{0}$ and ${\bm d}$ components.
For instance, it is possible to emerge the coexistence such as
\begin{align}
&d_{0}\text{ : an $s$-wave singlet with the even-$\omega$ dependence},
\notag\\
&d_{1}\text{ : a $p$-wave triplet with the even-$\omega$ dependence},
\notag\\
&d_{2}\text{ : a $p$-wave triplet with the even-$\omega$ dependence},
\notag\\
&d_{3}\text{ : an $s$-wave triplet with the odd-$\omega$ dependence},
\end{align}
purely by the symmetry consideration.


\begin{references}
\bibitem{RDS12} See for example, {\it Recent Developments in Superconductivity}, J. Phys. Soc. Jpn. (2012) {\bf 81}.
\bibitem{Berezinskii74} V.L. Berezinskii, Zh. Eksp. Teor. Fiz. Pis'ma Red. {\bf 20} 628 (1974) [JETP Lett. {\bf 20} 287 (1974)].
\bibitem{Kirkpatrick91} T.R. Kirkpatrick and D. Belitz, Phys. Rev. Lett. {\bf 66} 1533 (1991).
\bibitem{Belitz92} D. Belitz and T.R. Kirkpatrick, Phys. Rev. B {\bf 46} 8393 (1992).
\bibitem{Balatsky92} A.V. Balatsky, and E. Abrahams, Phys. Rev. B {\bf 45} 13125 (1992).
\bibitem{Abrahams93} E. Abrahams, A.V. Balatsky, J.R. Schrieffer, and P.B. Allen, Phys. Rev. B {\bf 47} 513 (1993).
\bibitem{Abrahams95} E. Abrahams, A.V. Balatsky, D.J. Scalapino, and J. R. Schrieffer, Phys. Rev. B {\bf 52} 1271 (1995).
\bibitem{Vojta99} M. Vojta, and E. Dagotto, Phys. Rev. B {\bf 59} 713 (1999).
\bibitem{Shigeta09} K. Shigeta, S. Onari, K. Yada, and Y. Tanaka, Phys. Rev. B {\bf 79} 174507 (2009).
\bibitem{Shigeta11} K. Shigeta, Y. Tanaka, K. Kuroki, S. Onari, and H. Aizawa, Phys. Rev. B {\bf 83} 140509  (2011).
\bibitem{Fuseya03} Y. Fuseya, H. Kohno, and K. Miyake, J. Phys. Soc. Jpn. {\bf 72} 2914(2003).
\bibitem{Kusunose11} H. Kusunose, Y. Fuseya, and K. Miyake, J. Phys. Soc. Jpn. {\bf 80} 044711 (2011).
\bibitem{Hotta09} T. Hotta, J. Phys. Soc. Jpn. {\bf 78} 123710 (2009).
\bibitem{Bergeret05} F.S. Bergeret, A.F. Volkov, and K.B. Efetov, Rev. Mod. Phys. {\bf 77} 1321 (2005).
\bibitem{Tanaka12} Y. Tanaka, M. Sato, and N. Nagaosa, J. Phys. Soc. Jpn. {\bf 81} 011013 (2012).
\bibitem{Bauer12} See for example, {\it Non-Centrosymmetric Superconductors}, Eds. E. Bauer, and M. Sigrist, (Springer, 2012).
\bibitem{Matsumoto12} M. Matsumoto, M. Koga, and H. Kusunose, J. Phys. Soc. Jpn. {\bf 81} 033702 (2012).
\bibitem{Aoki12} D. Aoki, and J. Flouquet, J. Phys. Soc. Jpn. {\bf 81} 011003 (2012).
\bibitem{Kusunose11a} H. Kusunose, Y. Fuseya, and K. Miyake, J. Phys. Soc. Jpn. {\bf 80} 054702 (2011).
\bibitem{Belitz99} D. Belitz, and T.R. Kirkpatrick, Phys. Rev. B {\bf 60} 3485 (1999).
\bibitem{Solenov09} D. Solenov, I. Martin, and D. Mozyrsky, Phys. Rev. B {\bf 79} 132502 (2009).
\bibitem{Luttinger60} J.M. Luttinger, and J.C. Ward, Phys. Rev. {\bf 118} 1417 (1960).
\bibitem{AGD} A.A. Abrikosov, L.P. Gorkov, and I.E. Dzyaloshinski, {\it Methods of Quantum Field Theory in Statistical Physics} (Dover, 1975).
\bibitem {Eliashberg60} G. M. Eliashberg, Sov. Phys. JETP {\bf 11} 696 (1960).
\bibitem{Potthoff03} M. Potthoff, Eur. Phys. J. B {\bf 32} 429 (2003).
\bibitem{notes1} The Zeeman splitting itself favors the triplet component with $S_{z}=\pm 1$, namely, ${\bm d}$ vector is perpendicular to the magnetic field. However, this does not mix with the singlet component, and the pure odd-frequency pairing is suppressed completely by the mass enhancement.
\bibitem{Carbotte90} J.P. Carbotte, Rev. Mod. Phys. {\bf 62} 1027 (1990).
\bibitem{Cappelluti07} E. Cappelluti and G.A. Ummarino, Phys. Rev. B {\bf 76} 104522 (2007).
\end{references}
\end{document}